\documentclass[12pt,preprint,notoc,nohyper]{crcsph} 

\usepackage{rotating}

\usepackage{epsfig,multicol,bbm}

\newcommand\fverb{\setbox\pippobox=\hbox\bgroup\verb}
\newcommand\fverbdo{\egroup\medskip\noindent%
            \fbox{\unhbox\pippobox}\ }
\newcommand\fverbit{\egroup\item[\fbox{\unhbox\pippobox}]}
\newbox\pippobox

\title{Conformal Ricci collineations of static spherically symmetric
spacetimes}

\author{Ugur Camci$^a$, Asghar Qadir$^b$ and
K. Saifullah$^c$ \footnote{\emph{On leave from:} Center for Advanced
Mathematics and Physics, National University of Sciences and
Technology, Rawalpindi, Pakistan, \emph{and} Department
of Mathematics, Quaid-i-Azam University, Islamabad, Pakistan.}\\

$^a$Department of Physics, Faculty of Arts and Sciences,
\d{C}anakkale Onsekiz Mart University, 17100 \d{C}anakkale, Turkey

$^b$Centre for Advanced Mathematics and Physics, National University
of Sciences and Technology, Rawalpindi, Pakistan

$^c$School of Mathematical
Sciences, Queen Mary, University of London, London, UK \\

Electronic address: \email{ucamci@comu.edu.tr},
\email{aqadirmath@yahoo.com}, \email{saifullah@qau.edu.pk}}

\preprint{}  

 \abstract{Conformal Ricci collineations of static spherically symmetric
spacetimes are studied. The general form of the vector fields
generating conformal Ricci collineations is found when the Ricci
tensor is non-degenerate, in which case the number of independent
conformal Ricci collineations is \emph{fifteen}; the maximum number
for 4-dimensional manifolds. In the degenerate case it is found that
the static spherically symmetric spacetimes always have an infinite
number of conformal Ricci collineations. Some examples are provided
which admit non-trivial conformal Ricci collineations, and perfect
fluid source of the matter.}

\dedicated{}

\begin{document}

\section{Introduction}

\label{INT}

Symmetries of Lagrangians are of great interest on account of
Noether's theorem, as they give information about the quantities
conserved under the motion. For the gravitational field this amounts
to the symmetries of the Ricci scalar density, $\sqrt{|g|}R$, where
$g$ is the determinant of the metric tensor, {\bf g}. Since the
curvature tensor, and hence the Ricci tensor, {\bf R}, and the Ricci
scalar are functions of the metric tensor and its first and second
derivatives, its differential symmetries are of fundamental
importance. These are called {\it isometries} or {\it Killing
vectors}. They satisfy the Killing equations
\begin{equation} \pounds_{\bf \xi} {\bf g} = 0. \label{isom}
\end{equation}
where $\pounds_{\bf \xi}$ signifies the Lie derivative operator
along the vector field ${\bf \xi}$. In component form these are
given by
\begin{equation} g_{ab,c} \xi^c + g_{ac} \xi_{,b}^c + g_{cb}
\xi^c_{,a} = 0. \label{isom1}
\end{equation}

The variation of the Lagrangian leads to the Einstein field
equations (EFEs) without the cosmological constant,
\begin{equation}R_{ab} - \frac{1}{2} R g_{ab} = \kappa T_{ab},
\end{equation}
where {\bf T} is the stress-energy tensor and gives the matter field
in the spacetime. As such, the symmetries of the matter tensor, {\bf
T} and of the Ricci tensor, are also of interest. Further, the
trace-free part of the curvature tensor, called the Weyl tensor {\bf
C}, gives the purely gravitational field and so its symmetries are
also of interest. These symmetries satisfy the equations obtained by
replacing the metric tensor in (\ref{isom}) by the relevant tensor
and are called {\it collineations} of that tensor. For the curvature
and Weyl tensors the component form becomes more complicated because
they have four indices.

The symmetries of the EFEs as differential equations are given by
replacing the right side of (\ref{isom}) by a constant times the
metric tensor\cite{stephani}, i.e.
\begin{equation}
\pounds_{\bf \xi}{\bf g} = 2 \sigma {\bf g}. \label{confm}
\end{equation}
These symmetries are called {\it homotheties} \cite{katzin}. If
$\sigma$ is not constant but an arbitrary function, the symmetries
are called {\it conformal isometries} or {\it conformal Killing
vectors} (CKVs). For other tensors they are called {\it conformal
collineations} of that tensor.

A procedure was developed to obtain all metrics that possessed some
minimal isometry group, e.g. SO(3) (spherical symmetry), and
classify them according to the higher symmetries they possess. The
general class of metrics with the given minimal symmetry would
remain where none of the conditions for the higher symmetry metrics
are met \cite{qz-a,BQZ,daud}. This procedure gave what was called
``the complete classification" of the spacetime with the minimal
symmetry by its isometries. It provided a list of all metrics that
possessed that symmetry and hence, in effect, all solutions of the
Einstein equations that possess the required symmetry. Since the
Einstein equations are highly non-linear coupled second order
partial differential equations, this procedure of obtaining
solutions was very useful. Because of the physical relevance of
homotheties, a complete classification by hometheties is also of
interest \cite{daud}. Further, the general solution of the conformal
Killing equations (\ref{confm}) in static spherically symmetric
spacetimes is given in Ref. 6. 
It is well-known that CKVs generate constants of motion along null
geodesics for massless particles. This property associates the
conformal symmetry with a well-defined, physically meaningful
conserved quantity.

It is of mathematical interest to look at classification by
collineations and conformal collineations of other tensors as well
and to investigate the relation between the various types of
collineations and conformal collineations. In this paper we have
limited our attention to 4-dimensional Lorentzian metrics of
signature ($+---$) and investigated the classification of static
spherically symmetric spacetimes by their conformal Ricci
collineations (also called \emph{Ricci inheritance collineations} by
Duggal\cite{duggal}). The manifold $M$ and the metric {\bf g} are
assumed to be smooth ($C^{\infty}$). Throughout this paper the usual
component notation in local charts will be used and a covariant
derivative with respect to the metric will be denoted by a semicolon
and a partial derivative by a comma.

We define \cite{tsamparlis1} a \emph{conformal Ricci collineation}
(CRC) ${\bf X}$ by
\begin{equation} \label{crc2}
R_{ab,c} X^c + R_{ac} X^c_{,b} + R_{cb} X^c_{,a} = 2 \psi R_{ab} ,
\end{equation}
where the conformal factor $\psi$ of conformal Ricci collineations
is a function of all the spacetime coordinates. It will be called a
\emph{homothetic Ricci collineation} (HRC) if $\psi$ is a constant,
and a \emph{Ricci collineation} (RC) if $\psi =0$. If we assume that
a vector field is simultaneously a CKV with conformal factor
$\sigma$ and a CRC with conformal factor $\psi$, then the relation
between the conformal factors $\sigma$ and $\psi$ is given by
\cite{ugur1}
\begin{equation}
\sigma_{;a b}= - \psi \left[ R_{ab} - \frac{1}{6} R \ g_{ab}
\label{cfactors} \right].
\end{equation}
When $\sigma_{;ab} =0$, i.e. ${\bf \xi}$ is a \emph{special} CKV,
then $\psi = 0$, i.e. ${\bf X}$ is an RC, and/or $R_{ab}
=\frac{1}{6} R \ g_{ab}$. The set of all CRCs on $M$ is a vector
space, but it may be infinite dimensional. If $R_{ab}$ is
non-degenerate, i.e. det$(R_{ab}) \neq 0$, then the Lie algebra of
CRCs is finite dimensional but if $R_{ab}$ is degenerate, it may not
be. If the Ricci tensor $R_{ab}$ is everywhere of rank $4$ then it
may be regarded as a metric on $M$ \cite{hall}. Thus, in the case of
a non-degenerate Ricci tensor, i.e. det$(R_{ab}) \neq 0$, we may
apply standard results on conformal symmetries to prove that the
maximal dimensions of the group of CRCs is 15 and this occurs if and
only if the \emph{Ricci tensor} regarded as a \emph{metric} is
conformally flat. Since the Ricci tensor arises naturally from the
Riemann curvature tensor (with components $R^a_{\ bcd}$ and where
$R_{ab} \equiv R^c_{\ acb}$) and hence from the connection, the
study of symmetries of the Ricci tensor $R_{ab}$ has a natural
geometrical significance \cite{tsamparlis2}-\cite{ugur4}.  The study
of the CRCs for the Friedmann-Robertson-Walker spacetimes has given
interesting results \cite{cb} and recently the relationship between
CRCs and CKVs for pp-waves has also been discussed in the literature
\cite{keane}. Furthermore, CRCs for the Einstein-Maxwell field
equations for non-null electromagnetic fields have also been studied
\cite{faridi}. It is worth noting that in the literature another
definition of CRCs also exists \cite{kuhnel} which is different from
the one given in this paper, and where a partial classification of
space-times carrying CRCs according to that definition has been
obtained.

    In this paper, we provide a complete classification of static
spherically symmetric spacetimes according to CRCs so that if we put
the conformal factor ($\psi$) equal to zero, the CRCs reduce to the
RCs obtained earlier \cite{qz} and if the conformal factor is taken
to be constant one gets the classification by HRCs \cite{bkk} where
$\psi$ is a constant. The plan of the paper is as follows. In the
next section, we give the CRC equations for spherically symmetric
static spacetimes. In Section \ref{degenerate} the CRC equations for
these spacetimes are solved when the Ricci tensor is degenerate,
while in Section \ref{non-degenerate} we obtain a general
classification for the non-degenerate case. It turns out that the
degenerate cases all have infinite dimensional Lie algebras of CRCs
and hence are not discussed beyond showing their existence. However,
for the non-degenerate cases we have looked, particulary, for
perfect fluid solutions with the desired symmetry. In the end, we
conclude with a brief summary and discussion of the results obtained
and give a few examples.

\section{The Conformal Ricci Collineation Equations}
\label{rics}
    The most general static spherically symmetric metrics, apart from the
Bertotti-Robinson classes, can be written as (\cite{qz-a}, and
references therein)
\begin{equation}\label{metric}
ds^2 =  e^{\nu (r)} dt^2 - e^{\lambda (r)} dr^2 - r^2 d\Omega^2,
\end{equation}
where $d\Omega^2 = d\theta^2 + \sin^2 \theta d\phi^2$. The
non-vanishing components of the Ricci tensor for this spacetime are
given by
\begin{eqnarray}
& & R_0 (r) \equiv R_{00} = \frac{1}{4} e^{\nu - \lambda} \left( 2
\nu'' + {\nu'}^2 - \nu' \lambda' + \frac{4}{r} \nu' \right),
\label{r0} \\& & R_1 (r) \equiv R_{11} = -\frac{1}{4} \left( 2 \nu''
+ {\nu'}^2 - \nu' \lambda' - \frac{4}{r} \lambda' \right),
\label{r1} \\& & R_2 (r) \equiv R_{22} = - \frac{1}{2} e^{-\lambda}
\left[ r (\nu' - \lambda') + 2 \right] + 1, \label{r2}\\& & R_3(r)
\equiv R_{33} = \sin^2 \theta R_2, \label{r2-r3}
\end{eqnarray}
where the prime denotes differentiation with respect to the radial
coordinate $r$. Then, the {\it Ricci tensor form} can be written as
\begin{equation}
ds^2_{Ric} \equiv R_{ab} dx^a dx^b = R_0 (r)  dt^2 + R_1 (r) dr^2 +
R_2 (r) d\Omega^2. \label{metric-ric}
\end{equation}
Using the EFEs with the perfect fluid which has the form $T_{ab} =
(\rho + p)u_a u_b - p g_{ab}$, where the fluid velocity is chosen as
$u_a = e^{\nu/2}$, one obtains
\begin{equation}
R_0 = \frac{e^{\nu}}{2} (\rho + 3 p), \quad R_1 =
\frac{e^{\lambda}}{2} (p - \rho), \quad R_2 = \frac{r^{2}}{2}(p -
\rho), \label{r012}
\end{equation}
where $\rho$ and $p$ are density and pressure of fluid,
respectively. Therefore, the Ricci tensor form becomes
\begin{equation}
2 ds^2_{Ric} = (\rho + 3p) e^{\nu} dt^2 - (\rho -p) \left[
e^{\lambda} dr^2 + r^2 d\Omega^2 \right],
\end{equation}
and is positive definite if and only if \emph{weak, strong} and
\emph{dominant energy conditions} which are respectively $\rho > 0,
\ \rho + 3 p > 0$ and $\rho >|p|$ are satisfied. For a perfect fluid
the energy conditions of a barotropic equation of state $p =
p(\rho)$ are given as
\begin{equation}
\rho >0, \quad 0 \leq p \leq \rho, \quad 0 \leq \frac{dp}{d\rho}
\leq 1 .
\end{equation}
The linear form of a barotropic equation of state is given by $p =
(\gamma -1)\rho$. If $\gamma =1, 4/3$ and $2$, then these imply
\emph{dust, incoherent radiation} and \emph{stiff matter}
respectively. For the static spherically symmetric spacetimes
(\ref{metric}), equation (\ref{crc2}) takes the form:
\begin{eqnarray}
& & R'_0 X^1 + 2 R_0 X^0_{,0}  = 2 R_0 \psi, \label{crc-a}
\\& & R'_1 X^1 + 2 R_1 X^1_{,1}  =  2 R_1 \psi, \label{crc-b}
\\& & R'_2 X^1 + 2 R_2 X^2_{,2} = 2 R_2 \psi, \label{crc-c}
\\& & R_2 \left( X^2_{,2} -\cot\theta X^2 - X^3_{,3} \right)= 0, \label{crc-d}
\\& & R_0 X^0_{,i} + R_i X^{i}_{,0} = 0, \label{crc-e}
\\ & & R_i X^{i}_{,j} + R_{j} X^{j}_{,i} = 0, \label{crc-f}
\end{eqnarray}
where $i,j = 1,2,3$, the summation convention is not assumed and $i
\neq j$ in (\ref{crc-f}). In this paper, we will take \emph{proper}
CRCs to denote CRCs that are not KVs or RCs. The general solution of
the CRC equations ((\ref{crc-a}) - (\ref{crc-f})) for the static
spherically symmetric spacetime (\ref{metric}) when $R_{ab}$ is
non-degenerate is obtained as
\begin{eqnarray}
X^0 &=& \frac{R_2}{R_0} \left[ \sin\theta \left( \dot{A}_1 \sin\phi
- \dot{A}_2 \cos\phi \right) + \dot{A}_3 \cos\theta \right] +
K(t,r), \label{x0}\\ X^1 &=& \frac{R_2}{R_1} \left[ \sin\theta
\left( A'_1 \sin\phi - A'_2 \cos\phi \right) + A'_3 \cos\theta
\right] + L(t,r), \label{x1} \\ X^2 &=& - \cos\theta \left[ A_1
\sin\phi - A_2 \cos\phi \right]+ A_3 \sin\theta + a_1 \sin\phi - a_2
\cos\phi, \label{x2} \\ X^3 &=& -csc\theta \left[ A_1 \cos\phi + A_2
\sin\phi \right]+ \cot\theta (a_1 \cos\phi + a_2 \sin\phi) + a_3,
\label{x3}
\end{eqnarray}
with the conformal function given by
\begin{eqnarray}
& & \psi = \left( \frac{R'_2}{2 R_2} A'_1 + A_1 \right) \sin\theta
\sin\phi - \left( \frac{R'_2}{2 R_2} A'_2 + A_2 \right) \sin\theta
\cos\phi  \nonumber \\ & & \qquad + \left( \frac{R'_2}{2 R_2} A'_3 +
A_3 \right) \cos\theta + \frac{R'_2}{2 R_2} L (t,r), \label{psi}
\end{eqnarray}
where $a_i$ are constants corresponding to the minimal KVs of
spherically symmetric spacetimes. Further, $A_i \equiv A_i (t,r), \
K(t,r)$ and $L(t,r)$ are subject to the following constraint
equations
\begin{eqnarray}
& & \left( \sqrt{\frac{R_1}{R_2}} L \right)' = 0, \label{cnst-a}
\\& & R_0 K' + R_1 \dot{L} = 0, \label{cnst-b} \\& & 2 \dot{K} +
\left( \frac{R'_0}{R_0} - \frac{R'_2}{R_2} \right) L = 0,
\label{cnst-c} \\& & \left( \sqrt{\frac{R_2}{R_0}} \dot{A}_i
\right)' = 0, \label{cnst-d}\\& & \ddot{A}_i + \frac{R_0}{2 R_1}
\left( \frac{R'_0}{R_0} - \frac{R'_2}{R_2} \right) A'_i -
\frac{R_0}{R_2} A_i = 0, \label{cnst-e}\\& & A''_i + \frac{1}{2}
\left( \frac{R'_2}{R_2} - \frac{R'_1}{R_1} \right) A'_i -
\frac{R_1}{R_2} A_i = 0, \label{cnst-f}
\end{eqnarray}
where dot represents the derivative with respect to time $t$, and
$i$ takes values $1,2,3$. Now, the problem of finding a CRC vector
field is reduced to solving the above constraint equations for the
corresponding case.
\section{Conformal Ricci Collineations for the Degenerate Ricci Tensor}
\label{degenerate} For the degenerate Ricci tensor, i.e. when
$det(R_{ab}) = 0$, we have the following possibilities: {\bf (a)}
all of the $R_a$ are zero; {\bf (b)} one of the $R_a$ is nonzero;
{\bf (c)} two of the $R_a$ are nonzero; {\bf (d)} three of the $R_a$
are nonzero. Then, for degenerate Ricci tensor cases, we use
directly the original CRC equations (\ref{crc-a})-(\ref{crc-f}).
Case {\bf (a)} corresponds to the vacuum (e.g. the Schwarzschild
spacetime) in which every vector is a CRC.

{\bf Case (b)}. We have the following subcases:
\begin{eqnarray}
&({\bf b.i})& R_0 \neq 0, \,\, R_i = 0, \,\, (i = 1,2,3);\,\, ({\bf
b.ii}) R_1 \neq 0, \,\, R_j = 0, \,\, (j = 0,2,3). \nonumber
\end{eqnarray}
In subcase (b.i), we find
\begin{eqnarray}
& X^0 = X^0(t), \quad X^1 = (\psi - \dot{X}^0)\frac{2 R_0}{R'_0},
\,\, X^{\alpha} = X^{\alpha} (x^a),\,\, \alpha = 2,3 \label{bi1}
\end{eqnarray}
where $R'_0 \neq 0$. If $R'_0 = 0$, i.e. $R_0 =c$ (a constant), then
we get
\begin{eqnarray}
& X^0 = \int{\psi dt} + a, \quad X^i = X^i (x^a), \label{bi2}
\end{eqnarray}
where $a$ is an integration constant. For this case, using the
conditions $R_i = 0$, we find that
\begin{eqnarray}
& e^{\lambda} = 1 + \frac{r}{2}(\nu - \lambda)', \quad R_0 =
\frac{e^{\nu-\lambda}}{r} (\nu + \lambda)', \\& 2 \nu'' +\nu'^2 -
\nu' \lambda' - \frac{4}{r} \lambda' = 0.
\end{eqnarray}
In this case we have the equation of state $p = \rho $ (stiff
matter), and the corresponding Lie algebra is infinite dimensional
due to the vector fields given in (\ref{bi1}) and (\ref{bi2}) have
arbitrary components. In subcase (b.ii), we obtain that
\begin{eqnarray}
& X^1 = \frac{1}{\sqrt{\mid R_1 \mid}} \left[ \int{\psi\,\sqrt{\mid
R_1 \mid} dr} + a_1 \right], \quad X^j = X^j (x^a), \label{bii}
\end{eqnarray}
where $a_1$ is a constant of integration. Then we have again
infinite dimensional Lie algebra for the latter subcase. For subcase
(b.ii), the solution of the conditions $R_j = 0$ give
\begin{eqnarray}
& e^{\lambda} = 1 + \frac{r}{2}(\nu - \lambda)', \quad e^{\lambda -
\nu} = c r^4 \nu'^2, \quad R_1 = \frac{1}{r}(\nu + \lambda)',
\end{eqnarray}
where $c$ is a non-zero constant. Further, for this case, perfect
fluid is not allowed, which means that when $R_1 \neq 0$ it must be
$R_2 \neq 0$ from (\ref{r012}) for a perfect fluid but this
contradicts with the condition $R_2 =0$ of this case.

{\bf Case (c)}. In this case, the possible subcases are given by
\begin{eqnarray}
&({\bf c.i})& R_p \neq 0, \,\, R_q = 0, \,\, (p = 0,1 \,\, {\rm and}
\,\, q= 2,3); \,\, ({\bf c.ii}) R_p = 0, \,\, R_q \neq 0. \nonumber
\end{eqnarray}
In the first subcase, using the transformations $d\bar{r} = \psi
\sqrt{\mid R_1 \mid} dr$, where $\bar{r}=\bar{r}(t,r)$, yields
\begin{equation}
X^0 = - \int{\frac{(\dot{\bar{r}}+\dot{f}) }{\psi R_0} d\bar{r}}
g(t), \quad X^ 1 = \frac{\bar{r} + f(t)}{\sqrt{\mid R_1 \mid}} ,
\quad X^q = X^q (x^a),
\end{equation}
where $f(t)$ and $g(t)$ are functions of integration, and the
conformal factor $\psi$ has the form
\begin{equation}
\psi = \frac{2 R_0}{(\bar{r} + f - 2R_0)} \left[ \int{
\left(\frac{\dot{\bar{r}} + \dot{f}}{\psi} \right)^{.}
\frac{d\bar{r}}{R_0}} - \dot{g}(t) \right].
\end{equation}
For this case, it follows from the conditions $R_q =0$ that
\begin{eqnarray}
& e^{\lambda} = 1 + \frac{r}{2}(\nu - \lambda)', \quad R_1 +
e^{\lambda -\nu} R_0 = \frac{1}{r}(\nu + \lambda)'.
\end{eqnarray}
In this case again perfect fluid is not allowed since $R_2 = 0$ but
$R_1 \neq 0$ which are contradictions for a perfect fluid
assumption. In subcase (c.ii), we have obtained
\begin{eqnarray}
& & X^0 = X^0 (x^a), \quad X^1 = \frac{2 R_2}{R'_2} \left[ \psi -
\sin\theta (A_1 \sin\phi - A_2 \cos\phi) - A_3 \cos\theta \right],
\nonumber \\& &  X^2 = -\cos\theta (A_1 \sin\phi - A_2 \cos\phi) +
A_3 \sin\theta + a_1 \sin\phi - a_2 \cos\phi, \nonumber \\& &  X^3 =
- \csc\theta (A_1 \cos\phi + A_2 \sin\phi ) +\cot\theta (a_1
\cos\phi + a_2 sin\phi ) + a_3,
\end{eqnarray}
where $A_i = b_i$ (constants), $R'_2 \neq 0$, and the scale factor
is given by
\begin{equation}
\psi = \frac{R'_2}{2 R_2} X^1 + \sin\theta ( A_1 \sin\phi - A_2
\cos\phi ) + A_3 \cos\theta.
\end{equation}
It is interesting to note that using the conditions $R_0 =R_1 =0$ of
this subcase, we have found the following metric
\begin{eqnarray}
& e^{\nu} = a \ e^{-\lambda}, \quad \nu^2 = -\frac{2b}{r}+ c,
\end{eqnarray}
where $a, b$ and $c$ are constants of integration, and $R_2 = 1
-\frac{e^{\nu}}{a} (1 + r \nu')$. Once again, in this case perfect
fluid is not allowed because of the condition that $R_1 = 0$ but $
R_2 \neq 0$.

{\bf Case (d)}. In this case, the possible subcases are as follows
\begin{eqnarray}
&({\bf d.i})& R_0 = 0, \,\, R_i \neq 0, \,\, (i = 1,2,3);\,\, ({\bf
d.ii}) R_1 = 0, \,\, R_j \neq 0, \,\, (j = 0,2,3). \nonumber
\end{eqnarray}
In subcase (d.i), we obtain that $X^0 = X^0 (x^a)$ and $X^i = X^i
(r,\theta,\phi)$ where the form of $X^i$ is the same as in equations
(\ref{x1}) - (\ref{x3}), and the constraint equations (\ref{cnst-a})
- (\ref{cnst-f}) for $A_i$ and $L$ have the following solution
\begin{equation}
A_i = b_i \cosh\bar{r} + d_i \sinh\bar{r}, \quad L = \ell \sqrt{
\frac{R_2}{R_1}},
\end{equation}
where $b_i, d_i$ and $\ell$ are constants. In this case, the
conformal factor is given by
\begin{eqnarray}
& & \psi = \left( \frac{R_{2,\bar{r}}}{2 R_2} A_{1,\bar{r}} + A_1
\right) \sin\theta \sin\phi - \left( \frac{R_{2,\bar{r}}}{2 R_2}
A_{2,\bar{r}} + A_2 \right) \sin\theta \cos\phi  \nonumber \\ & &
\qquad + \left( \frac{R_{2,\bar{r}}}{2 R_2} A_{3,\bar{r}} + A_3
\right) \cos\theta + \ell \frac{R_{2,\bar{r}}}{2 R_2},
\end{eqnarray}
where we have used the transformation $dr = (R_2/R_1)^{1/2}
d\bar{r}$. For this case we get the equation of state $\rho + 3p = 0
$. In subcase (d.ii), if $R_0 = R_2$, it follows from the constraint
equations (\ref{cnst-a}) - (\ref{cnst-f}) that $A_i$ and $K$ have
the following solutions
\begin{equation}
A_i = b_i \cosh\bar{r} + d_i \sinh\bar{r}, \quad K = \ell,
\end{equation}
where $b_i, d_i$ and $\ell$ are integration constants. Then, the
components of the CRC vector are obtained as
\begin{eqnarray}
& & X^0 = \sin\theta \left[ A'_1 \sin\phi - A'_2 \cos\phi \right] +
A'_3 \cos\theta + \ell, \nonumber \\& &  X^ 1 = \frac{2 R_0}{R'_0}
\left[ \psi - \sin\theta(A_1 \sin\phi - A_2 \cos\phi) - A_3
\cos\theta \right], \\& &  X^2 = -  \cos\theta (A_1 \sin\phi - A_2
\cos\phi) + A_3 \sin\theta + a_1 \sin\phi - a_2 \cos\phi, \nonumber
\\& & X^3 = - \csc \theta (A_1 \cos\phi + A_2 \sin\phi) + \cot\theta
(a_1 \cos\phi + a_2 \sin\phi) + a_3, \nonumber
\end{eqnarray}
where $R'_0 \neq 0$, and $\psi$ is arbitrary conformal factor, that
is, the component $X^1$ is arbitrary function of the coordinates and
so we have \emph{infinite} dimensional algebra. If $R_0 \neq R_2$,
we have the following CRCs
\begin{eqnarray}
& & X^0 = f(t), \nonumber \\& & X^ 1 = \frac{2 R_0 /R_2}{(R_0/R_2)'}
\left[ \sin\theta (A_1 \sin\phi - A_2 \cos\phi) + A_3 \cos\theta -
\dot{f}(t) \right] , \\& &  X^2 = - \cos\theta (A_1 \sin\phi - A_2
\cos\phi) + A_3 \sin\theta + a_1 \sin\phi - a_2 \cos\phi, \nonumber
\\& & X^3 = -\frac{1}{\sin\theta} (A_1 \cos\phi + A_2 \sin\phi) +
\cot\theta (a_1 \cos\phi + a_2 \sin\phi) + a_3, \nonumber
\end{eqnarray}
where $A_i = b_i$(constants), $(R_0/R_2)' \neq 0$,  $f(t)$ is an
integration function, and the conformal factor $\psi$ is given by
\begin{equation}
\psi =  \frac{R'_0/R_2}{(R_0/R_2)'}\left[ \sin\theta(A_1 \sin\phi -
A_2 \cos\phi) + A_3 \cos\theta  - \dot{f} \right] + \dot{f}.
\end{equation}
Thus, we again have an infinite dimensional algebra, and a perfect
fluid is not allowed due to the fact that $R_1 = 0$ but $R_2 \neq 0$
which give contradiction from (\ref{r012}) for a perfect fluid
assumption.

\section{Conformal Ricci Collineations for the Non-degenerate Ricci Tensor}
\label{non-degenerate} In this section, we consider the CRCs in
non-degenerate cases, i.e. det$(R_{ab}) \neq 0$, admitted by the
static spherically symmetric spacetimes. Thus, we have the following
possibilities and corresponding solutions of these possibilities.

{\bf Case (i)}. None of the $R'_a$ is zero. In this case, applying
the transformation $dr = \sqrt{R_2 /R_1} d\bar{r}$, the general
solutions of the constraint equations (\ref{cnst-a}) -
(\ref{cnst-f}) are obtained as
\begin{eqnarray}
& & A_i = \sqrt{\frac{R_0}{R_2}} \int{f_{i^{\pm}} (t) dt} + k_i h
(\bar{r}), \\& & K = - \frac{\sinh\bar{r}}{a} \sqrt{
\frac{R_2}{R_0}} \dot{f}_{4^{\pm}} (t) + \ell, \quad L =
\sqrt{\frac{R_2}{R_1}} f_{4^{\pm}} (t),
\end{eqnarray}
where $\ell$ is a constant, and $f_{i^{\pm}} (t), f_{4^{\pm}} (t)$
and $h (\bar{r})$ are as follows
\begin{eqnarray}
& & f_{i^{\pm}} (t) = \left\{
\begin{array}{l}
b_i \cosh(\alpha t ) + d_i \sinh( \alpha t), \quad
{\rm for} \, \alpha^2 > 0, \\
b_i \cos( |\alpha| t) + d_i \sin( |\alpha| t), \quad {\rm for} \,
\alpha^2 < 0,
\end{array}
\right.  \\& & f_{4^{\pm}} (t) = \left\{
\begin{array}{l}
m \cosh( \alpha t) + n \sinh( \alpha t), \quad   {\rm for} \,
\alpha^2 > 0, \\ m \cos( |\alpha| t) + n \sin( |\alpha| t), \quad
{\rm for} \, \alpha^2 < 0,
\end{array}
\right. \\ & & h (\bar{r}) = \cosh\bar{r} + \frac{a}{b}
\sinh\bar{r}, \\& & \alpha^2 = \left\{
\begin{array}{l}
a^2 - b^2, \quad {\rm for} \, \alpha^2 > 0, \\
b^2 - a^2, \quad {\rm for} \, \alpha^2 < 0,
\end{array}
\right.
\end{eqnarray}
where $R_0 = \left[ a \cosh\bar{r} + b \sinh\bar{r} \right]^2 R_2$,
$\alpha$ is a constant of separation, and $a, b$ are integration
constants. In addition to the parameters $a_i$ given in (\ref{x2})
and (\ref{x3}), which give the minimal KVs, we have also the
parameters $b_i, d_i, k_i, m, n$ and $\ell$ given throughout.
Therefore, it is seen that the number of CRCs is \emph{fifteen}. If
$\alpha^2 = 0$, we have the following solution of the constraint
equations
\begin{eqnarray}
& & A_i = e^{\beta \bar{r}} \left[ c_0 \left( b_i \ t + d_i \right)
+ k_i \right], \quad L = \sqrt{\frac{R_2}{R_1}} (m t + n)
\\& & K = \beta \ m \left( \frac{ e^{-2\epsilon \bar{r}}}{c^2_0}
- \frac{t^2}{2} \right) - \beta n t + \ell,
\end{eqnarray}
where $\beta = \pm 1$ and $R_0 = c^2_0 e^{2 \beta \bar{r}} R_2$,
which again gives \emph{fifteen} CRCs. In this case when $\alpha^2
\neq 0$ the Ricci tensor metric (\ref{metric-ric}) becomes
\begin{equation}
ds^2_{Ric} = R_2 (r) \left[ \left( a \cosh\bar{r} + b \sinh\bar{r}
\right)^2 dt^2 + d\bar{r}^2 +  d\Omega^2 \right].
\label{metric-ric2}
\end{equation}
It is found that the conformal tensor associated with the Ricci
tensor metric (\ref{metric-ric2}) vanishes. Similarly it is possible
to get the same result for the related Ricci tensor metric when
$\alpha^2 = 0$.

{\bf Case (ii)}. Three of the $R'_a$ are zero. In this case we have
the possibilities ({\bf ii.a}) $R'_0 \neq 0, \,\, R'_i = 0, \,\, (i
= 1,2,3)$ and ({\bf ii.b}) $R'_1 \neq 0, \,\, R'_j = 0, \,\, (j =
0,2,3)$. The results for subcase (ii.a) are given in Table 1. Using
the assumption of the perfect fluid, we find that
\begin{equation}
e^{\lambda} = c_1^{-2} c_2^2 r^{-2}, \quad R_0 = - (2 \rho + 3 c_2^2
r^{-2})e^{\nu}, \quad p = \rho + 2 c_2^2 r^{-2},
\end{equation}
where $R_1 = c_1^2, \, R_2 = c_2^2 $ ($c_1$ and $c_2$ are non-zero
constants). Now, let us take $c_1 = c_2 =1$. For this example, the
choice of $R_1 = R_2 = 1$ requires that $\bar{r} = r$, and the
spacetime admits \emph{fifteen} CRCs. In this case, we obtain the
following metric
\begin{equation}
ds^2 = e^{\frac{r^2}{2} \left( \frac{a\, r^2}{2} +1 \right)} \left[
{(b r)}^{-1} dt^2 - b r dr^2 \right] - r^2 d\Omega^2,
\label{metric-i}
\end{equation}
where $a$ and $b$ are non-zero arbitrary constants. For this case,
the Ricci tensor metric has the following form
\begin{equation}
ds^2_{Ric} = \left[ \alpha \cosh{r} + \beta \sinh{r} \right]^2 dt^2
+ d{r}^2 +  d\Omega^2 . \label{metric-ric3}
\end{equation}
It is found that all components of the conformal tensor associated
with this metric are zero.

In subcase (ii.b), the solutions of the constraint equations are
given by
\begin{equation}
A_i = 0, \quad K =  \frac{m}{c_0^2} \int{\sqrt{R_1}\ dr} + \ell,
\quad L =  \sqrt{R_1} ( m \ t + n),
\end{equation}
where $R_0 = c_0^2,\ R_2 =c_2^2$ and $m,n,$ and $\ell$ are
parameters, that is, the number of CRCs are \emph{six}. Then, we get
that $\psi = 0$, i.e. the CRCs reduce to the RCs. In this case, we
have obtained from EFEs with perfect fluid that
\begin{equation}
R_1 = c_2^2 r^{-2} e^{\lambda}, \quad p = \rho + 2 c_2^2
r^{-2},\quad \rho = -(c_0^2 e^{-\nu} + 3 c_2^2 r^{-2})/2.
\end{equation}
Here we take $c_0 = c_2 = 1$, i.e. $R_0 = R_2 = 1$ as an example of
this subcase. In this case, we obtain the following metric
\begin{equation}
ds^2 = r^a e^{\frac{b}{8} r^4}  \left[dt^2 - c^2 r^2 dr^2 \right] -
r^2 d\Omega^2, \label{metric-ii}
\end{equation}
where $a, b$ and $c$ are non-zero arbitrary constants. For this
case, CRCs reduce to the RCs, which are given by
\begin{eqnarray}
& & {\bf X}_{(1)} = \sin\phi \partial_{\theta} + \cos\phi \,
\cot\theta \partial_{\phi}, \,\, {\bf X}_{(2)} = \sin\phi
\partial_{\theta} + \cos\phi \, \cot\theta \partial_{\phi}, \nonumber
\\& & {\bf X}_{(3)} = \partial_{\phi}, \quad {\bf X}_{(4)} =
\partial_t,\quad {\bf X}_{(5)} = \partial_{\bar{r}}, \quad
{\bf X}_{(6)} = \bar{r} \partial_t - t \partial_{\bar{r}},
\end{eqnarray}
where we have used the rescaling $d\bar{r} = \sqrt{R_1} dr$.

{\bf Case (iii)}. Two of the $R'_a$ are zero. In this case, the
possible subcases are ({\bf iii.a}) $\ R'_p \neq 0, \,\, R'_q = 0,
\,\, (p = 0,1 \,\, {\rm and} \,\, q= 2,3)$ and ({\bf iii.b})$ R'_p =
0, \,\, R'_q \neq 0$. The results for these cases are summarized in
Table 1, in which we have used the functions $f_5 (t), h_{1^{\pm}},
h_{2^{\pm}}$ and $h_{3^{\pm}}$ defined by
\begin{eqnarray}
& & f_5 (t) = b_i t + d_i, \\ & & h_{1^{\pm}} (r) = \left\{
\begin{array}{l}
\cosh\left(\frac{c_1}{c_2} r \right) + \frac{a}{b} \sinh\left(
\frac{c_1}{c_2}r \right),
{\rm for} \,\, \alpha^2 > 0, \\
\cos\left( \frac{c_1}{c_2} r\right) - \frac{a}{b}
\sin\left(\frac{c_1}{c_2} r\right),  \ \ \ {\rm for} \,\, \alpha^2 <
0, \end{array} \right.  \\ & & h_{2^{\pm}} (\bar{r}) = \left\{
\begin{array}{l}
\cosh\left(\frac{\bar{r}}{c_2} \right) + \frac{a}{b} \sinh\left(
\frac{\bar{r}}{c_2} \right), \quad
{\rm for} \,\, \alpha^2 > 0, \\
\cos\left( \frac{\bar{r}}{c_2}\right) - \frac{a}{b}
\sin\left(\frac{\bar{r}}{c_2}\right), \quad \ \ \ {\rm for} \,\,
\alpha^2 < 0, \end{array} \right.  \\& & h_{3^{\pm}} (\bar{r}) =
\left\{
\begin{array}{l}
\cosh\left(c_1 \bar{r} \right) + \frac{a}{b} \sinh\left( c_1
\bar{r}\right), \quad {\rm for} \,\, \alpha^2 > 0, \\
\cos\left( c_1 \bar{r} \right) - \frac{a}{b} \sin\left( c_1 \bar{r}
\right), \quad \ \ \ {\rm for} \,\, \alpha^2 < 0,
\end{array}
\right.
\end{eqnarray}
where $c_1$ and $c_2$ are non-zero constants. For case (iii.a),
considering the assumption of the perfect fluid source in the EFEs,
it follows that $R_2 = c_2^2,$ and
\begin{equation}
R_0 = -(2 \rho + 3c_2^2 r^{-2})e^{\nu}, \quad R_1 = c_2^2 r^{-2}
e^{\lambda}, \quad p = \rho + 2 c_2^2 r^{-2}.
\end{equation}
In case (iii.b), we have found that $R_0 = c_0^2,\,\, R_1 = c_1^2$
and
\begin{equation}
R_2 = c_1^2 r^2 e^{-\lambda}, \quad p= \rho + 2 c_1^2 e^{-\lambda},
\quad \rho = -\frac{1}{2}(c_0^2 e^{-\nu} + 3 c_1^2 e^{-\lambda}),
\end{equation}
where $c_0$ is a non-zero constant. For the last subcase, if we take
$c_0 = c_1 = 1$, i.e. $R_0 = R_1 = 1$, then we obtain the following
metric
\begin{equation}
ds^2 =  \frac{a}{1+ \tan^2(r/2)}\left[dt^2 - a dr^2 \right] - r^2
d\Omega^2, \label{metric-iii}
\end{equation}
where $a$ is a non-zero arbitrary constant.

{\bf Case (iv)}. One of the $R'_a$ is zero. In this case, the
possible subcases are ({\bf iv.a})$ R'_0 = 0, \,\, R'_i \neq 0, \,\,
(i = 1,2,3)$ and ({\bf iv.b}) $\ R'_1 = 0, \,\, R'_j \neq 0, \,\, (j
= 0,2,3)$. For case (iv.a) one obtains similar results to case (i),
in which the only difference is the condition $R_0 = c_0^2$. In this
case, for the perfect fluid, we have found that $R_1 = (c_0^2
e^{-\nu} + 2 p)e^{\lambda}, \, R_2 = (c_0^2 e^{-\nu}+ 2 p)r^2$ and
$\rho + 3p = - 2 c_0^2 e^{-\nu}$. The solutions of the constraint
equations for the case (iv.b) are given in Table 1. For this case,
it follows from the EFEs with perfect fluid that $R_0 = (c_1^2
e^{-\lambda} -2p) e^{\nu}, \, R_1 = c_1^2, \, R_2 = c_1^2 r^2
e^{-\lambda}$, and $p = \rho + 2 c_1^2 e^{\lambda}$.

{\bf Case (v)}. All $R'_a$ are zero. In this case we have that $R_0
= c_0, \, R_1 = c_1, \, R_2 = c_2$,
\begin{equation}
A_i = 0, \quad K = \ell \frac{r}{c_0} + m, \quad L = -\ell
\frac{t}{c_1} + n,
\end{equation}
where $c_0, c_1$ and $c_2$ are non-zero constants, and $m, n$ and
$\ell$ are parameters. Then, it follows from this result that $\psi
= 0$, i.e. this case reduces to RCs. Further, in this case, using
the perfect fluid as a source of the EFEs we have
\begin{equation}
e^{\lambda} = \frac{c_1^2}{c_2^2} r^2, \quad p = \rho + 2 c_2^{2}
r^{-2}, \quad \rho = -\frac{1}{2}(3 c_2^2 r^{-2} + c_0^2 e^{-\nu}).
\end{equation}
The results for cases (ii) - (iv) are summarized in Table 1, and the
number of independent CKVs of the Ricci tensor form
(\ref{metric-ric}) \emph{or} CRCs of the original metric
(\ref{metric}) for all cases is \emph{fifteen}. In each of the cases
given above, the conformal function $\psi$ can be found from
(\ref{psi}) using the functions from Table 1.

\section{Conclusion}
\label{conc} In this paper, we have obtained a complete
classification of CRCs for static spherically symmetric spacetimes
not of Bertotti-Robinson type. We found that if the Ricci tensor is
degenerate, then the CRCs have infinite degrees of freedom. In the
non-degenerate case there are \emph{fifteen} CRCs. In the cases of
\emph{six} dimensional Lie algebras the conformal factor becomes
zero, therefore, they are RCs and not CRCs. In some cases we have
found that the conformal tensor associated with a \emph{metric}
whose coefficients are the components of the Ricci tensor, vanishes
identically. The significance of this finding needs to be
investigated. This will be done separately.

If $R'_2 = 0$ and $A_i = 0$, i.e. $\psi = 0$ (in which case CRCs
become RCs), then we have only \emph{six} RCs, which are given in
Ref. 14 (see equations (3.8) - (3.10) of this reference). This is an
example of the generality of our results. If $A_i = 0, R'_2 / 2 R_2
= a$ and $L = \ell$, then $\psi$ becomes a constant (namely $a
\ell$), and the constraint equations yield $K = m t + n, \ R_0 = c_0
e^{2 (a - m/\ell) r}, R_1 = c_1 R_2, \ R_2 = c_2 e^{2 a r}$. For
this solution, the number of parameters is \emph{five}. The detailed
analysis of the $\psi=$ constant case is given in Ref. 24.

To show that the classes of symmetries are non-empty it is necessary
to construct specific metrics that satisfy the constraints on the
Ricci tensor. Examples have been provided in different cases for
this purpose. We have then tried to construct perfect fluid
solutions for the various cases, providing perfect fluid spacetimes
with the desired CRC symmetries. However, in some cases of the
degenerate Ricci tensor a perfect fluid is not allowed, therefore,
one has to choose some other matter tensor for these cases.

The Bertotti-Robinson spacetime is a $V_2 \oplus S^2$. The conformal
symmetries of $S^2$ are very well known and the possible conformal
symmetries of a general $V_2$ are also well known, ranging from a
$0$ to $6$ dimensional Lie algebra. As such it is not worth
tabulating the classification here.

\acknowledgments

We would like to thank Prof. Graham Hall for useful discussions. Two
of us (UC and KS) are grateful to the Abdus Salam International
Centre for Theoretical Physics, Trieste, Italy, for travel grants
under its Net-53 to visit each other's institutions, and for local
hospitality to the National Centre for Physics and the Centre for
Advanced Mathematics and Physics, National University of Sciences
and Technology, Pakistan; and \d{C}anakkale Onsekiz Mart University,
\d{C}anakkale, Turkey. KS also acknowledges a research grant from
the Higher Education Commission of Pakistan.

\begin{sidewaystable}
\begin{center}
\begin{scriptsize}
\caption{\label{label}The functions $A_i (t,r), K(t,r), L(t,r)$ and
related transformations and constraints.}
{\begin{tabular}[t]{@{}llllll}
\hline Case&$A_i(t,r)$& $K(t,r)$& $L(t,r)$ &Transformations & Constraints\\
\hline (ii.a) & ${|R_0|}^{\frac{1}{2}} \int{f_{i^{\pm}} dt} + k_i
h_{1^{\pm}} $ & $- \frac{c_1 c_2 \sinh(c_1 r /c_2)}{a {|R_0|}^{1/2}
}\dot{f}_{4^{\pm}}  + \ell$& $ f_{4^{\pm}}$ & $---$
& $R_1 =\pm c_1^2, R_2 = \pm c_2^2,$ \\  & & & & & \qquad $\alpha^2 =\pm(a^2 -b^2)/c_2^2$ \\
& & & & & $R_0 = \left[ a \cosh(\frac{c_1}{c_2} r ) + b
\sinh(\frac{c_1}{c_2} r) \right]^2$
\\ & $ e^{\beta r} \left[ c_0 f_5 + k_i \right]$ & $\frac{m \beta}{2}
\left( \frac{c_2^2}{R_2} - t^2 \right) - \beta n t + \ell$ & $ m t +
n $  & $---$ & $R_0 = c_0^2 e^{2 \beta r}, \beta^2 = c_1^2/c_2^2 $\\
\hline (iii.a) &${|R_0|}^{\frac{1}{2}} \int{f_{i^{\pm}}  dt} + k_i
h_{2^{\pm}} $ & $- \frac{c_2 \sinh( \bar{r} /c_2)}{a {|R_0|}^{1/2}}
\dot{f}_{4^{\pm}}  + \ell$ &$ {|R_1|}^{-1/2}f_{4^{\pm}}  $ & $dr =
{|R_1|}^{-\frac{1}{2}} d\bar{r}$ & $R_2 = c_2^2,\, \alpha^2 =
\pm(a^2 -b^2)/c_2^2,$ \\ & & & & &  $ R_0 = \left[a
\cosh(\frac{\bar{r}}{c_2}) + b sinh(\frac{\bar{r}} {c_2}) \right]^2$
\\  & ${|R_0|}^{\frac{1}{2}} \int{f_{i^{\pm}}  dt} + k_i
h_{2^{\pm}}$ & $- \frac{c_2 \sin(\bar{r}/c_2)}{a {|R_0|}^{1/2}}
\dot{f}_{4^{\pm}}  + \ell$& $ {|R_1|}^{-1/2} f_{4^{\pm}} $ & {\rm
same as above} &$R_2 = - c_2^2,\, \alpha^2 = \pm (a^2 + b^2)/c_2^2$
\\ & & & & &$R_0 = \left[a \cos(\frac{\bar{r}}{c_2}) + b
sin(\frac{\bar{r}}{c_2}) \right]^2$
\\ & $ e^{\beta \bar{r}} \left[ c_0 f_5 (t) + k_i \right]$ &
$\frac{m \beta}{2} \left( \frac{c_2^2}{R_2} - t^2 \right) - \beta n
t + \ell$& $ {|R_1|}^{-1/2} (m t + n) $ & {\rm same as above}
& $R_0 = c_0^2 e^{2 \beta \bar{r}}, R_2 = c_2^2, \beta^2 = 1/c_2^2$ \\
\hline (iii.b)& ${|R_2|}^{\frac{1}{2}} \int{f_{i^{\pm}} dt} + k_i
h_{3^{\pm}} $ & $ - \frac{c_1}{a\, c_0^2} \sinh(c_1 \bar{r})
{|R_2|}^{\frac{1}{2}} \dot{f}_{4^{\pm}}  + \ell$ &
${|R_2|}^{\frac{1}{2}} f_{4^{\pm}}$ & $dr = {|R_2|}^{\frac{1}{2}}
d\bar{r}$ & $R_0 = c_0^2, R_1 = c_1^2, \alpha^2 = c_0^2 (a^2 -b^2)$
\\ & & & & & $R_2 = \left[ a\cosh(c_1 \bar{r}) + b \sinh(c_1
\bar{r}) \right]^{-2}$
\\ & ${|R_2|}^{\frac{1}{2}} \int{f_{i^{\pm}} dt} + k_i h_{3^{\pm}}$
& $ \frac{c_1}{a\, c_0^2} \sin(c_1 \bar{r}) {|R_2|}^{\frac{1}{2}}
\dot{f}_{4^{\pm}} + \ell$ & ${|R_2|}^{\frac{1}{2}} f_{4^{\pm}}$ &
{\rm same as above} & $R_0 = \pm c_0^2, R_1 = - c_1^2,$ \\  & & & &
& \qquad $\alpha^2 = c_0^2 (a^2 +b^2)$ \\ & & & & & $R_2 = \left[
a\cos(c_1 \bar{r}) + b \sin(c_1 \bar{r}) \right]^{-2}$ \\ &
$e^{-\beta \bar{r}} \left[ c_2^{-1} f_5 + k_i \right]$ &$ \frac{m
\beta}{2} \left( \frac{R_2}{c_0^2} - t^2 \right) + \beta n t + \ell$
& ${|R_2|}^{\frac{1}{2}} (m t + n) $ & {\rm same as above} & $ R_2 =
c_2^2 e^{2 \beta \bar{r}}, \ \beta = c_1^2$ \\ \hline (iv.b) &
${|\frac{R_0}{R_2}|}^{\frac{1}{2}}\int{f_{i^{\pm}} dt} + k_i
h_{3^{\pm}} $ & $-\frac{c_1}{a} \sinh(c_1 \bar{r})
{|\frac{R_2}{R_0}|}^{\frac{1}{2}} \dot{f}_{4^{\pm}}  + \ell $ &
${|R_2|}^{\frac{1}{2}} f_{4^{\pm}}$ & $dr = {|R_2|}^{\frac{1}{2}}
d\bar{r}$ & $R_1 = c_1^2, \ \alpha^2 = a^2 -b^2$\\& & & & & $R_0 =
\left[ a \cosh(c_1 \bar{r}) + b \sinh(c_1 \bar{r}) \right]^2 R_2 $\\
& ${|\frac{R_0}{R_2}|}^{\frac{1}{2}}\int{f_{i^{\pm}} dt} + k_i
h_{3^{\pm}} $ & $\frac{c_1}{a} \sin(c_1 \bar{r})
{|\frac{R_2}{R_0}|}^{\frac{1}{2}} \dot{f}_{4^{\pm}} + \ell $ &
${|R_2|}^{\frac{1}{2}} f_{4^{\pm}}$ & {\rm same as above} & $R_1 =
-c_1^2, \ \alpha^2 = a^2 + b^2$\\& & & & & $R_0 = \left[ a \cos(c_1
\bar{r}) + b \sin(c_1 \bar{r}) \right]^2 R_2 $\\ & $e^{\beta
\bar{r}} \left[ c_0 f_5 + k_i \right]$ & $ \frac{m \beta}{2} \ell
\left( \frac{R_2}{R_0} - t^2 \right) - \beta n t + \ell$ &
${|R_2|}^{\frac{1}{2}}(m t + n)$ & {\rm same as above} & $R_0 = c_0
e^{2 \beta \bar{r}} R_2, \beta = c_1^2 $
\\ \hline
\end{tabular}}
\end{scriptsize}
\end{center}
\end{sidewaystable}

\end{document}